\documentclass[reprint, superscriptaddress, amsmath, amssymb, aps, pre, floatfix]{revtex4-2}

\usepackage{cancel}
\usepackage{graphicx}% Include figure files
\usepackage{dcolumn}% Align table columns on decimal point
\usepackage{bm}% bold math
\usepackage[normalem]{ulem}	% for text strikethrough;  use \sout{text}

\usepackage[T1]{fontenc}
\usepackage{mathptmx}
\usepackage{etoolbox}

\usepackage{enumerate}

\usepackage{cancel}
\usepackage{soul, xcolor}

\usepackage{physics}
\usepackage{xr}
\usepackage[colorlinks=true,citecolor=black,linkcolor=black,urlcolor=blue]{hyperref}
\newcolumntype{P}[1]{>{\centering\arraybackslash}p{#1}}

\newcommand{\kT}{k_{\rm B}T}
\newcommand{\kB}{k_{\rm B}}
\newcommand{\tauR}{\tau_{\rm r}}

\DeclareMathOperator {\e}{e}						% e for exponential should be roman

\newcommand\redsout{\bgroup\markoverwith{\textcolor{red}{\rule[0.5ex]{2pt}{0.4pt}}}\ULon}

\begin{document}

\title{Harnessing higher-dimensional fluctuations in an information engine}

\author{Antonio Patr\'on Castro}%
\email{apatronc@sfu.ca}
\affiliation{Department of Physics, Simon Fraser University, Burnaby, British Columbia V5A 1S6, Canada}
\author{John Bechhoefer}%
\email{johnb@sfu.ca}
\affiliation{Department of Physics, Simon Fraser University, Burnaby, British Columbia V5A 1S6, Canada}
\author{David A.\ Sivak}%
\email{dsivak@sfu.ca}
\affiliation{Department of Physics, Simon Fraser University, Burnaby, British Columbia V5A 1S6, Canada}

\begin{abstract}
    We study the optimal performance of an information engine consisting of an overdamped Brownian bead confined in a controllable, $d$-dimensional harmonic trap and additionally subjected to gravity. The trap's center is updated dynamically via a feedback protocol designed such that no external work is done by the trap on the bead, while maximizing the extraction of gravitational potential energy and achieving directed motion. We show that performance improves when thermal fluctuations in directions perpendicular to gravity are harnessed. This improvement arises from feedback cooling of these transverse degrees of freedom, along which all heat is extracted{. Strikingly, engines based on a single transverse degree of freedom already outperform engines based solely on {vertical} ($z$) measurements.} This engine design modularizes the functions of harnessing fluctuations and storing free energy, drawing a close analogy to the Szilard engine.
\end{abstract}

\maketitle
%\tableofcontents
%\newpage

Over a century ago, Maxwell proposed his famous thought experiment, suggesting that information about a system’s microscopic dynamics could be used to extract useful energy without any work input -- seemingly violating the second law of thermodynamics~\cite{knot_tait_1911}. In the 1930s, Szilard refined this idea by introducing the first concrete model of what is now known as an \textit{information engine}~\cite{szilard_entropie_1929,szilard_maxwell_2003}: a cyclic device that exploits thermal fluctuations by applying feedback, thereby extracting heat from a thermal bath. Its operation is reconciled with the second law by Landauer’s principle: processing and erasing information about a system's dynamics has a minimum cost~\cite{landauer_irreversibility_1961,bennett_thermodynamics_1982}.

Recent advances in technology and stochastic thermodynamics~\cite{sekimoto_kinetic_1997,sekimoto_stochastic_2010,seifert_stochastic_2012,vandenbroeck_ensemble_2015,seifert_stochastic_2025} have enabled the experimental realization of modern information engines~\cite{toyabe_information_2010,camati_entropy_2016,koski_refrigerator_2015,cottet_quantum_2017,masuyama_conversion_2018,koski_szilard_2014,chida_power_2017,admon_experimental_2018,goerlich_experimental_2025,baldovin_optimal_2024}. This capability has been used to test Landauer’s principle and quantify the cost of information processing~\cite{berut_verification_2012,jun_test_2014,koski_mutual_2014,hong_test_2016,ciliberto_landauers_2021,dago_information_2021,archambault_information_2025}, cool nanoparticles to millikelvin temperatures~\cite{li_fundamental_2012,gieseler_feedback_2012,tebbenjohanns_cold_2019}, and confirm the limits of the second law~\cite{barros_probabilistic_2024,paneru_reaching_2020,klinger_universal_2025}. Inspired by the ideas of Maxwell, Szilard, and Landauer, researchers have even constructed molecular-scale information engines, demonstrating that synthetic cyclic molecular machines can indeed leverage fluctuations to power their operation~\cite{serreli_molecular_2007}. More recently, it has been suggested that many of the molecular motors operating within living cells 
(such as ATP synthase or kinesin)
may also work as information engines, harnessing energy from the noisy cellular environment~\cite{leighton_efficiencies_2023,leighton_arbitrage_2024,leighton_flow_2025,tsuruyama_rna_2023,parrondo_thermodynamics_2015}.

In previous work~\cite{du_buisson_performance_2024}, Saha, et al.\ designed and experimentally realized an information engine that, in addition to extracting energy from a thermal bath, stored such energy in a gravitational potential by raising a weight. Their engine consisted of a micron-scale bead in water, harmonically confined via optical tweezers~\cite{lucero_fluctuation_2021,saha_optical_2021}. The trap center was raised upon measuring a favorable `up' fluctuation of the bead (parallel to gravity). Subsequent studies explored performance for Bayesian inference of the bead’s position under noisy measurements~\cite{saha_bayesian_2022} and for nonequilibrium active noise~\cite{saha_nonequilibrium_2023,paneru_2022_colossal,paneru_optimal_2018}.

To date, information engines have been essentially one-dimensional, exploiting fluctuations along a single degree of freedom. Can harnessing fluctuations along additional degrees of freedom further improve performance?  In this work, we generalize a theoretical study of the experimental engine of \cite{saha_maximizing_2021} to $d$ dimensions and find striking increases in the rate of energy extraction and related measures of performance. We show that the performance enhancement results from feedback cooling of thermal fluctuations along the transverse degrees of freedom, thereby extracting \textit{all} available heat, since all such fluctuations are favorable. We demonstrate that feedback on the transverse degrees of freedom alone can produce high output power, by analyzing an engine variant in which we do not measure the vertical $z$ component. This engine design separates the essential functions of harnessing fluctuations and storing free energy, capturing a core feature of the original Szilard engine. These results highlight the potential of higher-dimensional fluctuations as a valuable resource in the design of information engines { and underscore the importance of choosing which degrees of freedom to measure and their impact in overall engine performance.}

\begin{figure}
  \centering
  \includegraphics[width=0.8\columnwidth,height=0.56\columnwidth]{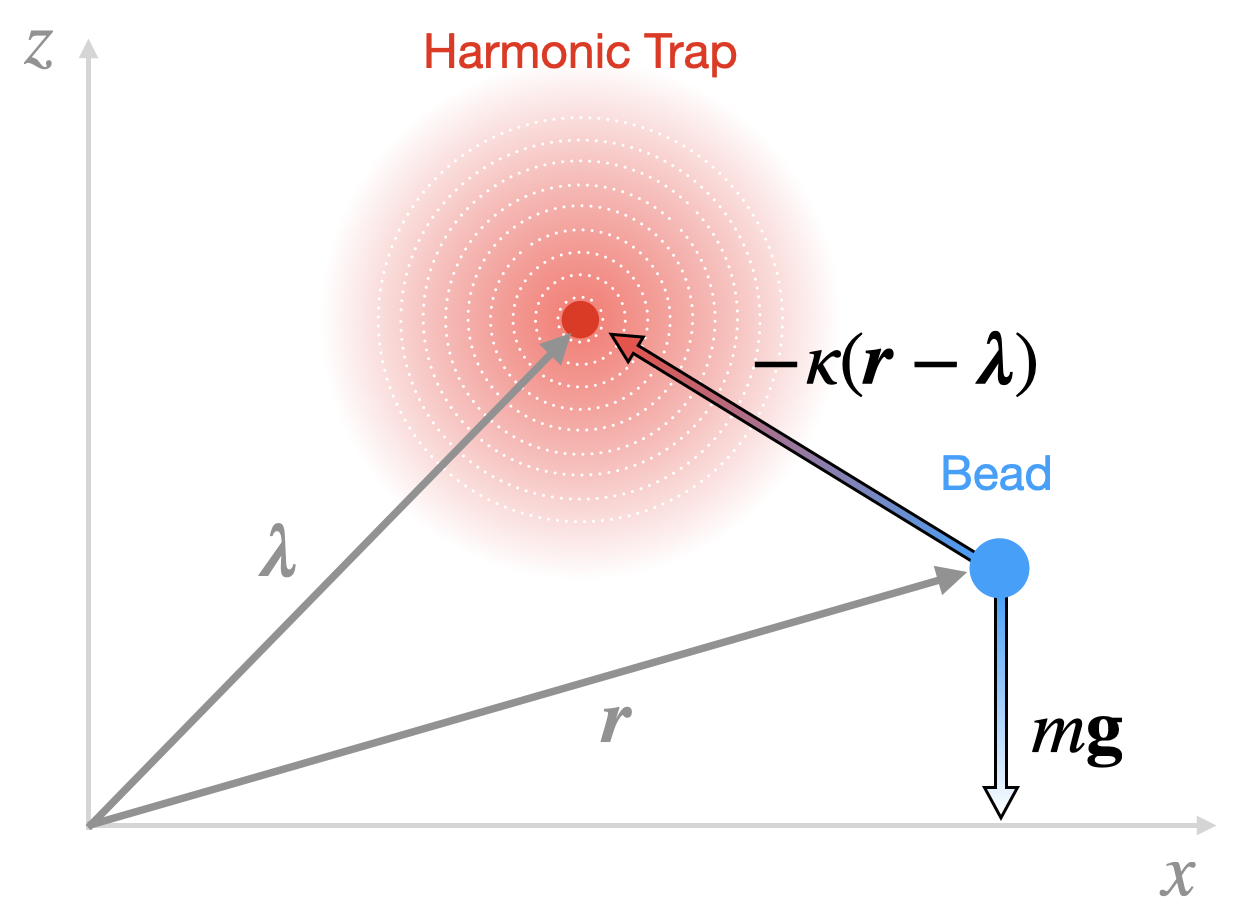}
  \caption{
  Schematic for $d\!=\!2$: a Brownian bead (blue dot) at position $\boldsymbol{r} = (x,z)$ experiences the gravitational force in the $z$-direction and the action of a confining optical trap (red dot and area) centered at $\boldsymbol{\lambda}$.
  }
  \label{fig:sketch}
\end{figure}

\textit{Multidimensional information engine}\label{sec:model}---Consider an optically trapped bead in $d$ spatial dimensions, whose dynamics are governed by the overdamped Langevin equation
\begin{equation}\label{eq:langevin}
    \gamma \, \dot{\boldsymbol{r}}'(t')=-\kappa[\boldsymbol{r}'(t')-\boldsymbol{\lambda}'(t')] - m  \boldsymbol{g} + \sqrt{2\kT\gamma} \ \boldsymbol{\xi}(t') \ ,
\end{equation}
for bead position $\boldsymbol{r}'(t')$ at time $t'$, isotropic harmonic trap center $\boldsymbol{\lambda}'(t')$, trap stiffness $\kappa$, bead mass $m$ relative to the surrounding fluid, gravitational acceleration $\boldsymbol{g}$, friction coefficient $\gamma$, Boltzmann constant $\kB$, and thermal-bath temperature $T$. $\boldsymbol{\xi}(t')$ is a vector of Gaussian white-noise fluctuations of zero mean and correlations
\begin{equation}
\left<\xi^{(i)}(t_1')\, \xi^{(j)}(t_2')\right> = \delta_{ij} \, \delta(t_1'-t_2'), \quad \forall \, i,j=1,...,d \ .
\end{equation}
We nondimensionalize the problem by scaling lengths with the equilibrium standard deviation $\sigma_{\text{eq}} \equiv \sqrt{\kT/\kappa}$ of the bead's position and time with the relaxation time $\tauR \equiv \kappa / \gamma$. This yields the dimensionless variables $t \equiv t' / \tau_{\rm r}, \boldsymbol{r}(t) \equiv \boldsymbol{r}'(t')/\sigma_{\text{eq}}$ and $\boldsymbol{\lambda}(t)\equiv \boldsymbol{\lambda}'(t')/\sigma_{\text{eq}}$, and
the dimensionless overdamped Langevin equation
\begin{equation}
    \dot{\boldsymbol{r}}(t)=-[\boldsymbol{r}(t)-\boldsymbol{\lambda}(t)] - \boldsymbol{\delta_{\rm g}} + \sqrt{2} \ \boldsymbol{\xi}(t) \ ,
\end{equation}
where $\boldsymbol{\delta_{\rm g}}\equiv m \boldsymbol{g}/\kappa \sigma_{\text{eq}}$ is a dimensionless vector quantifying the gravitational force relative to the amplitude of thermal fluctuations. Figure~\ref{fig:sketch} illustrates the system for $d\!=\!2$.

The total, time-dependent potential for the bead results from the combination of the optical trap and the gravitational force,
\begin{align}\label{eq:potential}
    V(\boldsymbol{r}, \boldsymbol{\lambda}(t)) &= \frac{1}{2}|\boldsymbol{r}-\boldsymbol{\lambda}(t)|^2 + \boldsymbol{r}^{\top} \boldsymbol{\delta_{\rm g}} \ . 
\end{align}
The bead's position is measured at discrete time intervals of duration $t_{\rm s}$, and the feedback on the trap position is applied immediately. Integrating Eq.~\eqref{eq:langevin} over one time step provides the discrete-time dynamics~\cite{saha_maximizing_2021,kloeden_numerical_1992},
\begin{equation}\label{eq:langevin-discrete}
    \boldsymbol{r}_{n+1} = \e^{-t_{\rm s}}\boldsymbol{r}_n + (1-\e^{-t_{\rm s}})(\boldsymbol{\lambda}_{n^+} - \boldsymbol{\delta_{\rm g}}) + \sqrt{1-\e^{-2t_{\rm s}}} \ \boldsymbol{\nu}_n \ ,
\end{equation}
for timestep number $n$ and vector $\boldsymbol{\nu}_n$ of Gaussian random variables with zero mean and correlations
\begin{equation}
\left<\nu_{n\phantom{'}}^{(i)} \, \nu_{n'\phantom{'}}^{(j)}\right> = \delta_{ij} \, \delta_{nn'}, \quad \forall \, i,j=1,...,d \ .
\end{equation}
The subscript $n^+$ indicates that $\boldsymbol{\lambda}$ is updated \textit{after} measuring $\boldsymbol{r}$. To simplify analysis and aid physical interpretation, we choose an orthogonal coordinate system such that $\boldsymbol{r} = (x^{(1)}, x^{(2)}, \dots, x^{(d-1)}, z)$ and $\boldsymbol{\lambda} = (\lambda^{(1)}, \lambda^{(2)}, \dots, \lambda^{(d-1)}, \lambda^{(z)})$, for $z$ the degree of freedom parallel to the gravitational force. 
%The discrete Langevin equation~\eqref{eq:langevin-discrete} becomes
%\begin{subequations}\label{eq:langevin-coords}
%\begin{align}
%z_{n+1} &= \e^{-t_{\rm s}} z_n + (1-\e^{-t_{\rm s}})(\lambda_{n^+}^{(z)}-\delta_{\rm g}) + \sqrt{1-\e^{-2t_{\rm s}}}\, \nu_{n}^{(z)} \\
%x_{n+1}^{(i)} &= \e^{-t_{\rm s}} x_{n}^{(i)} + (1-\e^{-t_{\rm s}})\lambda_{n^+}^{(i)} + \sqrt{1-\e^{-2t_{\rm s}}}\, \nu_{n}^{(i)} \ ,
%\end{align}
%\end{subequations}
%with $i = 1, 2, \dots, d-1$. Equations~\eqref{eq:langevin-coords} show that 
With this choice of coordinates, $z$ is influenced by both the gravitational force and the harmonic trap, whereas the remaining degrees of freedom are only driven by the harmonic trap. Throughout, $\{x^{(i)}\}$ and $\{\lambda^{(i)}\}$ are the transverse components of the bead and trap center, respectively, which live on the orthogonal $(d-1)$-dimensional space $\mathbb{R}_{d-1}$.

After updating bead position $\boldsymbol{r}_n$ to $\boldsymbol{r}_{n+1}$, the trap position $\boldsymbol{\lambda}$ is updated according to a feedback algorithm, chosen (following \cite{saha_maximizing_2021}) to optimize two performance metrics. The first metric is the net output power 
\begin{equation}\label{eq:output-power}
    P_{\text{net}} = f_{\rm s}\left(\left<\Delta F\right> - \left<W\right> \right) \ ,
\end{equation}
for sampling frequency $f_{\rm s} \equiv 1/t_{\rm s}$, and per-measurement average stored equilibrium free energy $\left<\Delta F\right>$ and work $\left<W\right>$ done on the bead. The averages are taken over the system probability distribution at steady state. Here, work is the instantaneous change in the total potential when the trap center $\boldsymbol{\lambda}$ is updated, 
\begin{subequations}
\begin{align}\label{eq:work}
    W_{n+1} &\equiv V(\boldsymbol{r}_{n+1},\boldsymbol{\lambda}_{n^++1}) - V(\boldsymbol{r}_{n+1},\boldsymbol{\lambda}_{n^+}) \\
    &= \frac{1}{2}|\boldsymbol{r}_{n+1}-\boldsymbol{\lambda}_{n^++1}|^2 - \frac{1}{2}|\boldsymbol{r}_{n+1}-\boldsymbol{\lambda}_{n^+}|^2 \ .
\end{align}
\end{subequations}
By convention, work is positive if energy flows into the system from the harmonic trap. Similarly, the average stored equilibrium free energy is
\begin{equation}\label{eq:free-energy-main}
    \left<\Delta F\right> = \delta_{\rm g} \left< \lambda_{n^++1}^{(z)}-\lambda_{n^+}^{(z)}\right> \ .
\end{equation}

From the first law of thermodynamics for the bead at steady state, the net output power $P_{\text{net}}$ equals the rate of heat extraction from the environment~\cite{du_buisson_performance_2024}. This follows from the assumption that no energy is exchanged between the bead and the controller, such that all stored energy is in the form of gravitational free energy. We focus on pure information engines, where exactly no work is done by each trap movement ($W_{n+1}=0$). As a result, these engines store useful free energy by exploiting thermal fluctuations alone: heat is continuously extracted from the thermal bath without any external work input. 

\begin{figure}
    \centering
    \includegraphics[width=0.8\columnwidth,height=0.56\columnwidth]{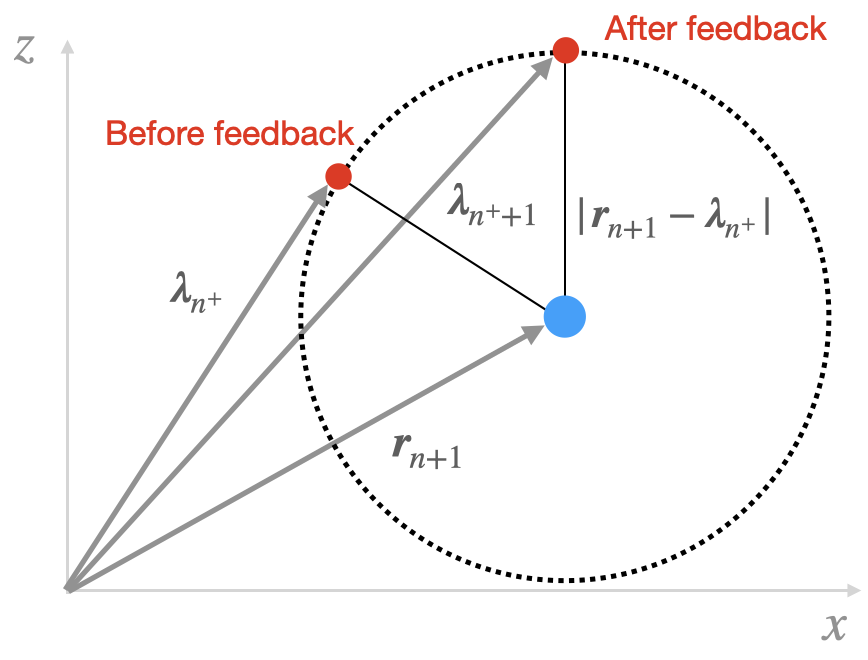}
    \caption{
    Schematic of the zero-work condition and the optimal feedback rule maximizing free-energy storage for $d\!=\!2$. The dashed circle shows possible updated positions $\boldsymbol{\lambda}_{n^++1}$ for the trap center, given $\boldsymbol{r}_{n+1}$ and $\boldsymbol{\lambda}_{n^+}$, for which no work is exerted on the bead. The updated trap center lies at the top of a hypersphere centered at $\boldsymbol{r}_{n+1}$ with radius $|\boldsymbol{r}_{n+1}-\boldsymbol{\lambda}_{n^+}|$. 
    %Thus the $z$-component of $\boldsymbol{\lambda}_{n^++1}$---and hence, the average stored equilibrium free energy~\eqref{eq:free-energy-main}---is maximized. 
    }
    \label{fig:feedback}
\end{figure}

The feedback algorithm is then chosen such that (i)  the harmonic trap does no work on the system and (ii) the stored free energy is maximized [\cite{supp_mat}~I], giving
\begin{equation}\label{eq:feedback-rule}
    \boldsymbol{\lambda}_{n^++1} = \boldsymbol{r}_{n+1} + |\boldsymbol{r}_{n+1}-\boldsymbol{\lambda}_{n^+}|\hat{\boldsymbol{z}} \ ,
\end{equation}
for unit vector $\hat{\boldsymbol{z}}$. The zero-work condition $W_{n+1}=0$ constrains $\boldsymbol{\lambda}_{n^++1}$ to a hypersphere centered at $\boldsymbol{r}_{n+1}$ with radius $|\boldsymbol{r}_{n+1}-\boldsymbol{\lambda}_{n^+}|$, and the free-energy maximization further restricts $\boldsymbol{\lambda}_{n^++1}$ to the top of this hypersphere. Thus a down fluctuation in $z$ does not prevent harnessing a simultaneous lateral fluctuation to lift the equilibrium trap position. Figure~\ref{fig:feedback} illustrates both the zero-work condition and the optimal feedback rule~\eqref{eq:feedback-rule} for a $d\!=\!2$ pure information engine. 

Resolving the vertical and transverse components gives
\begin{subequations}\label{eq:feedback-rule-components}
    \begin{align}
        \lambda_{n^++1}^{(z)} &= z_{n+1} + \sqrt{(z_{n+1}-\lambda_{n^+}^{(z)})^2+\sum_{j=1}^{d-1}\left(\Delta x_{n+1}^{(j)}\right)^2} \label{eq:feedback-rule-components-1} \\
        \lambda_{n^++1}^{(i)} &= x_{n+1}^{(i)}, \quad i=1,2,...,d-1 \ , \label{eq:feedback-rule-components-2}
    \end{align}
\end{subequations}
with bead update $\Delta x_{n+1}^{(j)}\equiv x_{n+1}^{(j)}-x_{n}^{(j)} = x_{n+1}^{(j)}-\lambda_{n^+}^{(j)}$ equaling the bead position relative to the trap center. The trap's $z$-component stores gravitational free energy, while the transverse components track the bead's transverse position (so transverse fluctuations are isotropic), effectively enacting a \emph{feedback cooling} protocol that maximizes heat extraction in Brownian engines~\cite{lee_experimentally-achieved_2018}. Any gain in potential energy due to transverse fluctuations is transferred to stored free energy during the subsequent trap update. 

The second performance metric is the long-time average $z$-velocity, $\left<v_z\right>$, which for a pure information engine is proportional to the output power~\cite{saha_maximizing_2021}:
\begin{equation}\label{eq:average-vel}
    P_{\text{net}} = \delta_{\rm g} \left<v_z\right> \ .
\end{equation}
This metric is directly accessible in experiments and quantifies the information engine's ability to generate directed motion against the load from an external force.

\begin{figure}
  \centering
  \includegraphics[width=1\columnwidth,height=0.55\columnwidth]{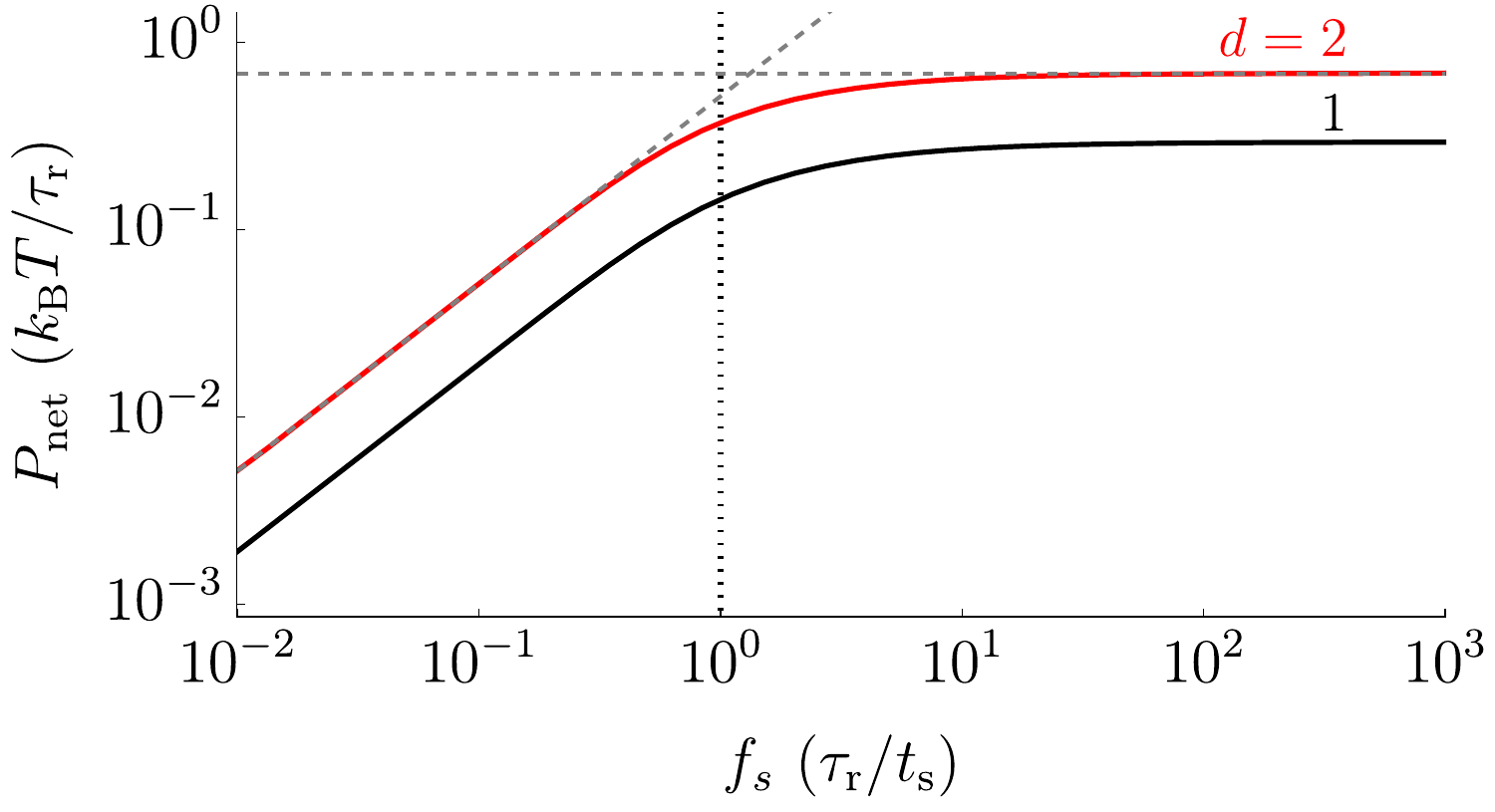}
  \caption{ 
  Information-engine output power $P_{\text{net}}$ as a function of sampling frequency $f_{\rm s}$, for $d\!=\!2$ (red) and $d\!=\!1$ (black). Solid curves: semi-analytic steady-state calculations [\cite{supp_mat}~IV and \cite{saha_maximizing_2021}]. Dashed gray lines: analytic results in the low-sampling-frequency limit [\cite{supp_mat}~III] and the high-sampling-frequency limit~\eqref{eq:max-output-power}. The dotted vertical line indicates $f_{\rm s}=1$. We set $\delta_{\rm g} = 0.8$, the value that approximately maximizes the output power for $d\!=\!1$~\cite{saha_maximizing_2021}.
  }
  \label{fig:output-power}
\end{figure}

\textit{Optimal performance}---We first examine the dependence of performance on the sampling frequency $f_{\rm s}$ for fixed $\delta_{\rm g} = 0.8$, chosen because it was previously shown to (approximately) maximize the output power for $d\!= \!1$~\cite{saha_maximizing_2021}. In Fig.~\ref{fig:output-power}, we compare the performance of a $d\!=\!2$ engine with the previously characterized $d\!=\!1$ engine (Fig.~3A in~\cite{saha_maximizing_2021}). The two engines exhibit similar behavior. At high sampling frequencies, the output power and heat extraction saturate. In this regime, the trap-center dynamics approximately follow a Langevin-like equation, producing maximum output power
\begin{equation}\label{eq:max-output-power}
    P_{\text{net}}^{\text{HF}}(\delta_{\rm g})= (d-1)\delta_{\rm g} \frac{\mathcal{Z}_{d-1}(\delta_{\rm g})}{\mathcal{Z}_{d}(\delta_{\rm g})} \ ,
\end{equation}
for partition function
\begin{equation}\label{eq:partition-fun}
    \mathcal{Z}_{d}(\delta_{\rm g}) \equiv \int_0^{\infty}\dd{L} \, L^{d-1} \, \e^{-L^2/2+\delta_{\rm g} L} \ ,
\end{equation}
expressable in terms of hypergeometric functions [\cite{supp_mat}~II]. Equation~\eqref{eq:max-output-power} holds even for $d \to 1$, recovering the $d\!=\!1$ expression for the output power from \cite{saha_maximizing_2021}.

At low sampling frequencies $f_{\rm s}$, ratchet events occur with frequency proportional to $f_{\rm s}$, as the bead’s position equilibrates between measurements. Consequently, the average stored equilibrium free energy in Eq.~\eqref{eq:free-energy-main} reaches a constant value, so the output power is linear in $f_{\rm s}$ [\cite{supp_mat}~III]. The transition between the two limiting behaviors takes place around $f_{\rm s}=1$, where the sampling time $t_{\rm s} \approx \tauR$. Despite the similar trends, the $d\!=\!2$ engine nearly doubles the output power attained by its $d\!=\!1$ counterpart.

The comparison between the performances of the $d\!=\!2$ and $d\!=\!1$ information engines already suggests that fluctuations in degrees of freedom perpendicular to gravity can be exploited to enhance heat extraction. In \cite{supp_mat}~V, we investigate whether it is better to exploit all such transverse fluctuations or {to wait for { rare} large ones which give a higher increase in stored free energy}. To address this, we introduce a modified feedback rule, in which
%\begin{equation}\label{eq:feedback-rule-2}
%    \boldsymbol{\lambda}_{n^++1} = 
%    \begin{cases}
%      \boldsymbol{r}_{n+1} + %|\boldsymbol{r}_{n+1}-%\boldsymbol{\lambda}_{n^+}|\hat{\boldsymbol{z}} %\ , & \! \sum_{j=1}^{d-1}
%      \left(
%      \Delta x_{n+1}^{(j)}
%      \right)^2 
%      \geq R^2 \\
%      \boldsymbol{\lambda}_{n^+} \ , & %\text{otherwise} \ .
%    \end{cases}
%\end{equation}
%That is, 
the trap center is updated only when, at the time of measurement, the bead lies outside a cylinder of radius $R$.
%{---accounting for the amplitude of transverse fluctuations}
{ The threshold $R$ setting the minimum fluctuation size is centered on the trap position} and aligned with the $z$-axis. 
%We study the dependence of the output power on $R$ in the high-sampling-frequency limit, for which we have already shown that the performance is optimal. In this limit, the output power can be interpreted as the ratio of extracted work to the first passage time required for the bead to reach the cylinder. Two competing effects arise as a function of the threshold $R$. On one hand, the extracted work per ratchet event increases with $R$: the larger the threshold, the greater the radius at which the zero-work condition is met, and thus the more work is extracted. On the other hand, the mean first-passage time also increases with $R$, since the bead must travel a greater distance before triggering a trap update~\cite{bray_persistence_2013,archambault_information_2025}[\cite{supp_mat}~\ref{app:app:threshold}]. Figure~\ref{fig:output-power-threshold} shows the output power in the high-sampling-frequency limit as a function of $R$. 
We show that the output power decreases monotonically with $R$, maximized at $R = 0$. Therefore, the optimal strategy exploits \textit{all} available transverse fluctuations{, regardless of their 
size}.

%\begin{figure} 
%    \centering
%    \includegraphics[width=0.86\columnwidth,height=0.6\columnwidth]{DtMF2.png}
%    \caption{
%    Semi-analytic computation of output power $P^{\text{HF}}_{\text{net}}$ [\cite{supp_mat}~\ref{app:app:threshold}] in the high-sampling-frequency limit ($f_{\rm s} \to \infty$) as a function of the threshold $R$ in the transverse degree of freedom, for $d\!=\!2$ and $\delta_{\rm g}\!=\!0.8$. Inset: schematic showing bead (blue), trap center (red) and surrounding cylinder of radius $R$.}
%    \label{fig:output-power-threshold}
%\end{figure}

Having shown that continuous sampling and continuous ratcheting optimize performance, we study the dependence of the output power and the vertical velocity on the dimensionless force $\delta_{\rm g}$. For $d\!=\!1$, the optimal output power at $\delta_{\rm g}^* \! \approx \! 0.8$ arises from two competing factors: On the one hand, as we increase $\delta_{\rm g}$ from $0$, the potential energy to extract increases. On the other hand, the magnitude of the opposing force also increases, making significant up fluctuations unlikely. This competition leads to maximum output power at an intermediate $\delta_{\rm g}$. 

\begin{figure} 
    \centering
    \includegraphics[width=1\columnwidth,height=1.125\columnwidth]{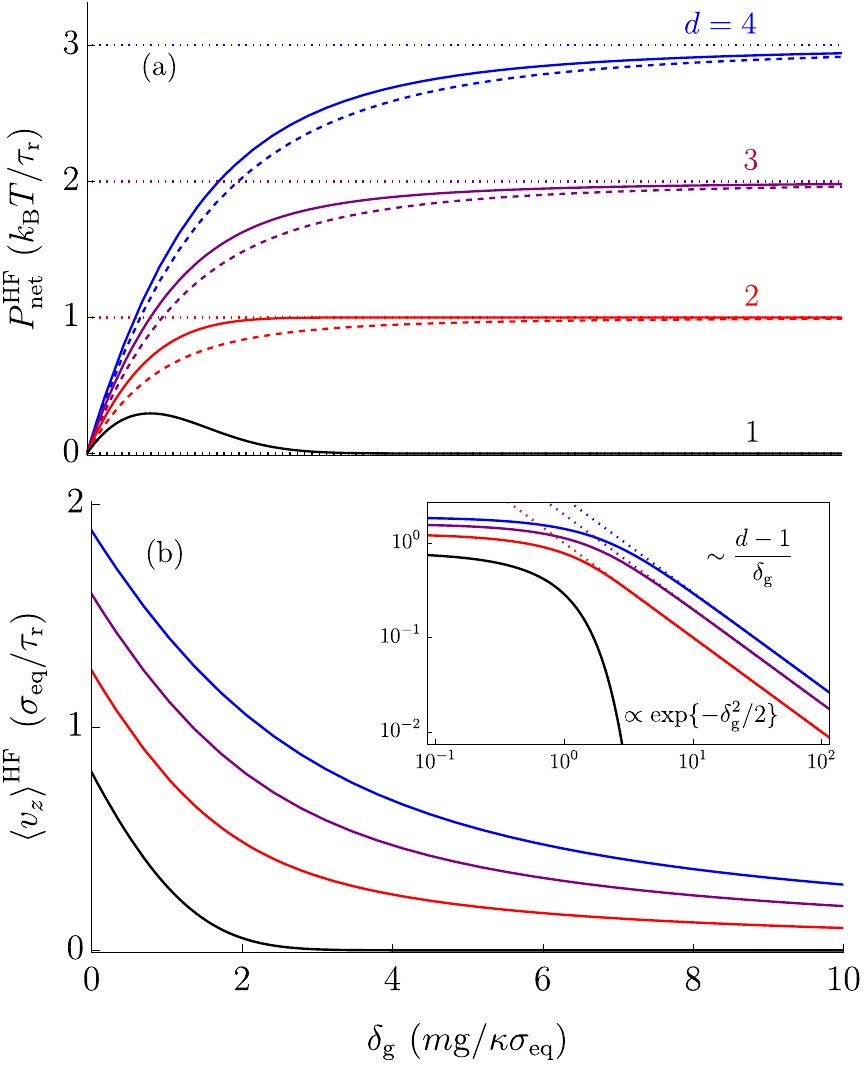}
    \caption{
    Output power (a) and vertical velocity (b) as functions of the scaled gravitational force $\delta_{\rm g}$, in the high-sampling frequency limit, for different dimensions. Solid curves: analytic results~\eqref{eq:max-output-power} [\cite{supp_mat}~II]. Dashed curves: partial information engine from Eq.~\eqref{eq:power-averaged}. Dotted lines: asymptotic limit $d-1$ for $\delta_{\rm g} \to \infty$.
    %, as given by Eq.~\eqref{eq:asympt}. 
    Inset in (b): log-log plot highlighting the algebraic decay for large $\delta_{\rm g}$, whose analytic form is represented by dotted lines.
    }
    \label{fig:power-vel-HF-high-d}
\end{figure}

Surprisingly, the nonmonotonic dependence on $\delta_{\rm g}$ is not seen in higher dimensions. Figure~\ref{fig:power-vel-HF-high-d} shows the output power and the vertical velocity as functions of the opposing gravitational force $\delta_{\rm g}$, for different dimensions. The output power is maximized in the $\delta_{\rm g} \to \infty$ limit for $d>1$, asymptoting to $d-1$, which implies that the velocity decays to zero algebraically as $\delta_{\rm g}$ increases ($\sim 1/\delta_{\rm g}$), in contrast to the exponential $\text{exp}\{-\delta_{\rm g}^2/2\}$ decay in one dimension~\cite{saha_maximizing_2021}. Approximating the partition function~\eqref{eq:partition-fun} for $\delta_{\rm g} \to \infty$ using Laplace's method~\cite{bender_advanced_1978} 
%gives the asymptotic limit
%\begin{equation}
%    \mathcal{Z}_d(\delta_{\rm g}) \sim %\sqrt{2\pi} \, \delta_{\rm g}^d \e^{\delta_{\rm %g}^2/2} \ ,
%\end{equation}
%resulting in
gives $P_{\text{net}}^{\text{HF}}(\delta_{\rm g}) \sim d-1$.
%\begin{align}\label{eq:asympt}
%    P_{\text{net}}^{\text{HF}}(\delta_{\rm g}) &\sim d-1 \quad \delta_g \gg 1 \ .
%\end{align}

The fact that the output power asymptotically approaches $d\!-\!1$ (in dimensionless units) for large $\delta_{\rm g}$ has striking implications. The output power is bounded above by the value attained under feedback cooling---where the trap follows the bead after each measurement, thereby extracting all the thermal fluctuations in potential energy. For feedback cooling, the power extracted by each independent degree of freedom is $1$~\cite{lee_experimentally-achieved_2018}, and hence the overall power for $d\!-\!1$ feedback-cooled degrees of freedom equals $d-1$. Our findings suggest that our information engine extracts the maximum possible power from the $d\!-\!1$ transverse degrees of freedom, while the $z$-component contributes no significant heat extraction, since in this limit useful up-fluctuations against gravity become very unlikely. This interpretation is supported by the explicit component-wise form of the feedback rule in Eq.~\eqref{eq:feedback-rule}: the $z$ degree of freedom adjusts to satisfy the zero-work condition, whereas the transverse degrees of freedom follow the feedback-cooling protocol, thereby maximizing heat extraction along those directions.

\textit{Ignoring vertical fluctuations}\label{sec:average}---In the previous section, we showed that, in the $\delta_{\rm g} \gg 1$ limit, the output power of a $d$-dimensional information engine is equal to feedback cooling of the $d\!-\!1$ degrees of freedom perpendicular to the gravitational force. The $z$-component fluctuations, parallel to gravity, do not contribute to the engine's performance. This observation raises a natural question: Is it necessary to measure the $z$-component in this regime, or is it sufficient to measure only the transverse components? To address this, we consider a \textit{partial} information engine, in which the feedback rule depends only on measurements of the $d\!-\!1$ transverse degrees of freedom, ignoring the $z$-component of the bead's position. For clarity, we refer to the previously studied system as the \textit{complete} information engine.

Since we do not measure the $z$-component, the zero-work condition $W_{n+1}=0$ cannot be rigorously fulfilled at each time step. Instead, we design a feedback rule for a pure information engine such that the average work $\left<W_{n+1}\right>$ is zero. This dictates a feedback rule for $\lambda_{n^++1}^{(z)}$ of the form
\begin{align}\label{eq:feedback-rule-average}
    \lambda_{n^++1}^{(z)} =& \ \lambda_{n^+}^{(z)} + \left< z_{n+1} - \lambda_{n^+}^{(z)}\right> \nonumber \\
    &+ \sqrt{\left< z_{n+1} - \lambda_{n^+}^{(z)}\right>^2 + \sum_{j=1}^{d-1}\left(\Delta x_{n+1}^{(j)}\right)^2} \ ,
\end{align}
with $\lambda_{n^++1}^{(i)}$ following Eq.~\eqref{eq:feedback-rule-components-2}. The averages are taken over the steady-state distribution for the  relative displacements $z_{n+1}-\lambda_{n^+}^{(z)}$. 
%Since we do not know this steady-state distribution \emph{a priori}, we consider the general family of feedback rules
%parametrized by $\omega$, 
%\begin{equation}\label{eq:feedback-rule-parametrized}
%    \lambda_{n^++1}^{(z)} = \lambda_{n^+}^{(z)} + \omega + \sqrt{\omega^2 + \sum_{j=1}^{d-1}\left(\Delta x_{n+1}^{(j)}\right)^2} \ ,
%\end{equation}
%then compute the average $\left<z_{n+1}-\lambda_{n^+}^{(z)}\right>$ and set $\omega$ to match the feedback rule~\eqref{eq:feedback-rule-average}. In this context, $\omega$ constitutes the (initially unknown) average change of the $z$-component of the trap-bead displacement.
In \cite{supp_mat}~VI, 
%we derive a self-consistency equation for $\omega$, leading to an 
we derive an analytical expression for the output power in the limit of high sampling frequencies,
\begin{equation}\label{eq:power-averaged}
    P_{\text{net}}^{\text{HF}} (\delta_{\rm g}) = \frac{2(d-1)}{1+\sqrt{1 + 4(d-1)/\delta_{\rm g}^2}} \ .
\end{equation}

Figure~\ref{fig:power-vel-HF-high-d}a shows that the complete and partial information engines attain the same output power for $\delta_{\rm g} \to \infty$, consistent with our previous result that, in that limit, the $z$ fluctuations do not contribute to free-energy extraction. For intermediate $\delta_{\rm g}$, Fig.~\ref{fig:power-vel-HF-high-d}a shows that measuring $z$ has some small advantage, permitting the harnessing of some $z$ fluctuations. But even in the simplest scenario, measuring a single transverse degree of freedom ($x$ for the $d\!=\!2$ partial information engine) increases output power remarkably compared to measuring only $z$ as in the original $d\!=\!1$ engine.

\textit{Conclusions}\label{sec:conclusions}---We have studied the theoretical performance of an information engine that harnesses thermal fluctuations in a $d$-dimensional overdamped system to achieve directed motion against gravity, thus converting heat from the thermal bath into stored gravitational potential energy. Systematically optimizing the engine's output power and velocity, we have shown that our $d\! >\! 1$-dimensional engine significantly outperforms the previously studied $d\!=\!1$ engine~\cite{saha_maximizing_2021,lucero_fluctuation_2021}. %{The enhancement of stored free energy per dimension can be viewed in terms of the increase in the available phase-space volume that follows the symmetry-breaking event of measuring the bead’s position, as pointed out in Ref.~\cite{roldan_universal_2014}.}
{The increase in stored free energy per dimension can be viewed as a decrease in the available phase-space volume 
%that follows 
following
the symmetry-breaking event of measuring the bead’s position~\cite{roldan_universal_2014}.}

The mechanism behind this enhancement is intimately related to feedback cooling: the $d\!-\!1$ transverse degrees of freedom each extract $\kT/\tauR$---the maximum amount of heat available per dimension---whereas heat extraction in the $z$-component is limited by gravity. Indeed, in the $\delta_{\rm g} \to \infty$ limit, the output power saturates to $d\!-\!1$ (in units of $\kT/\tauR$), implying that operation is sustained only by the transverse degrees of freedom. This is further confirmed by the partial information engine considered in Eq.~\eqref{eq:feedback-rule-average}: even without measurements of the bead's $z$-component, the engine still achieves the same limiting performance for large $\delta_{\rm g}$, with minor deviations from the complete engine for intermediate values of $\delta_{\rm g}$. These results suggest that choosing which degrees of freedom to measure may play a dominant role in the overall engine performance. This is similar to the problem of identifying suitable reaction coordinates for efficient enhanced sampling of free-energy landscapes in complex molecular systems~\cite{ma_reaction_2025,louwerse_information_2022}. Thus, our work provides a design principle that could guide the next-generation of nanoscale energy-harvesting devices~\cite{kolchinsky_maximising_2025,lucente_optimal_2025,olsen_harness_2025}. 

In this work, we have focused on maximizing engine performance, without considering the information costs that result from nearly continuous measurement of the bead's position. An important direction for future work involves the experimental realization of the $d\!=\!2$ and $d\!=\!3$ engines based on nanoscale Brownian objects confined in optical traps{, which are straightforward extensions of existing $d\!=\!1$ information engines~\cite{lee_experimentally-achieved_2018, saha_maximizing_2021}.} Further research could also consider the measurement and information-processing costs of the measuring devices and uncertainties in position and time measurements{. A full accounting of all entropy production requires including these information-processing costs}~\cite{toyabe_information_2010,ribezzi-crivellari_large_2019,admon_experimental_2018,paneru_lossless_2018,still_thermodynamic_2020}. The high-sampling-frequency limit (where we measure much faster than the system's characteristic dynamical time scales) pays large information costs for diminishing benefit. The partial information engine from Eq.~\eqref{eq:feedback-rule-average} is thus more efficient, gathering and processing less information to attain similar heat extraction.

More broadly, it would be interesting to explore the role of higher-dimensional fluctuations in different information engines involving more complex physical mechanisms. Promising directions include underdamped dynamics~\cite{dago_information_2021,dago_dynamics_2022,dago_adiabatic_2023,archambault_inertial_2024,sanders_optimal_2024,sanders_minimal_2025}, where applying feedback to velocity measurements improves engine control~\cite{bechhoefer_control_2021,kim_entropy_2004}; { 
multicomponent molecular motors, where energy and information flow between the system's different components to transduce
free energy \cite{leighton_efficiencies_2023,leighton_arbitrage_2024,leighton_flow_2025,tsuruyama_rna_2023,parrondo_thermodynamics_2015}; }and active noise, where the nonequilibrium bath serves as an additional source of fluctuations~\cite{malgaretti_szilard_2022,cocconi_efficiency_2024,datta_second_2022,lee_brownian_2020,garcia_optimal_2025,paneru_2022_colossal}, giving information engines a potential advantage over conventional heat engines.

%\appendix

\emph{Acknowledgments}.---
We thank Matthew Leighton (Yale Physics) and Johan du Buisson (Tubingen Theoretical Physics) for fruitful feedback on the manuscript. 
%This work was supported by a Natural Sciences and Engineering Research Council of Canada (NSERC) Discovery Grant and Discovery Accelerator Supplement RGPIN-2020-04950 (DAS), 
This work was supported by Natural Sciences and Engineering Research Council of Canada (NSERC) Discovery Grants ({JB}, DAS) and by Discovery Accelerator Supplement RGPIN-2020-04950 (DAS), 
an NSERC Alliance International Collaboration Grant ALLRP-2023-585940 (DAS), and a Tier-II Canada Research Chair CRC-2020-00098 (DAS). This research was enabled in part by support provided by BC DRI Group and the Digital Research Alliance of Canada (www.alliancecan.ca).

\emph{Data availability}.---The codes that support the findings of
this Letter are openly available~\cite{apatron_github_2025}.

\bibliography{Info2D_bibliography}

@misc{supp_mat,
  title        = {See {Supplemental} {Material} for details of the derivations and numerical methods, which includes {Refs}.~\cite{saha_maximizing_2021,kloeden_numerical_1992,bender_advanced_1978}},
}

@misc{apatron_github_2025,
  author = {Patr{\'o}n Castro, A.},
  year   = {2025},
  note   = {\href{https://github.com/apatroncastro/Harnessing-higher-dimensional-fluctuations-in-an-information-engine}{GitHub repository}}
}

@article{roldan_universal_2014,
	title = {Universal features in the energetics of symmetry breaking},
	volume = {10},
	issn = {1745-2473, 1745-2481},
	url = {https://www.nature.com/articles/nphys2940},
	doi = {10.1038/nphys2940},
	number = {6},
	urldate = {2025-10-09},
	journal = {Nature Physics},
	author = {Rold\'an, \'E. and Mart\'inez, I. A. and Parrondo, J. M. R. and Petrov, D.},
	month = jun,
	year = {2014},
	pages = {457--461},
}

@article{klinger_universal_2025,
  title = {Universal energy-speed-accuracy trade-offs in driven nonequilibrium systems},
  author = {Klinger, J\'er\'emie and Rotskoff, Grant M.},
  journal = {Phys. Rev. E},
  volume = {111},
  issue = {1},
  pages = {014114},
  numpages = {7},
  year = {2025},
  month = {Jan},
  publisher = {American Physical Society},
  doi = {10.1103/PhysRevE.111.014114},
  url = {https://link.aps.org/doi/10.1103/PhysRevE.111.014114}
}

@article{olsen_harness_2025,
  author  = {Olsen, Kristian Stølevik and Goerlich, R{\'e}mi and Roichman, Yael and L{\"o}wen, Hartmut},
  title   = {Harnessing non‐equilibrium forces to optimize work extraction},
  journal = {Nat. Commun.},
  volume  = {16},
  number  = {1},
  pages   = {11031},
  year    = {2025},
  doi     = {10.1038/s41467-025-67114-8},
  url     = {https://doi.org/10.1038/s41467-025-67114-8}
}

@misc{lucente_optimal_2025,
      title={Optimal Control of an Electromechanical Energy Harvester}, 
      author={Dario Lucente and Alessandro Manacorda and Andrea Plati and Alessandro Sarracino and Marco Baldovin},
      year={2025},
      eprint={2501.07735},
      archivePrefix={arXiv},
      primaryClass={cond-mat.stat-mech},
      url={https://arxiv.org/abs/2501.07735}, 
}

@Article{kolchinsky_maximising_2025,
AUTHOR = {Kolchinsky, Artemy and Marvian, Iman and Gokler, Can and Liu, Zi-Wen and Shor, Peter and Shtanko, Oles and Thompson, Kevin and Wolpert, David and Lloyd, Seth},
TITLE = {Maximizing Free Energy Gain},
JOURNAL = {Entropy},
VOLUME = {27},
YEAR = {2025},
NUMBER = {1},
ARTICLE-NUMBER = {91},
URL = {https://www.mdpi.com/1099-4300/27/1/91},
PubMedID = {39851711},
ISSN = {1099-4300},
DOI = {10.3390/e27010091}
}

@article{baldovin_optimal_2024,
  title = {Optimal Control of Levitated Nanoparticles through Finite-Stiffness Confinement},
  author = {Baldovin, Marco and Ben Yedder, Ines and Plata, Carlos A. and Raynal, Damien and Rondin, Lo\"{\i}c and Trizac, Emmanuel and Prados, Antonio},
  journal = {Phys. Rev. Lett.},
  volume = {135},
  issue = {9},
  pages = {097102},
  numpages = {7},
  year = {2025},
  month = {Aug},
  publisher = {American Physical Society},
  doi = {10.1103/pss4-b69w},
  url = {https://link.aps.org/doi/10.1103/pss4-b69w}
}

@article{goerlich_experimental_2025,
doi = {10.1209/0295-5075/adbb17},
url = {https://dx.doi.org/10.1209/0295-5075/adbb17},
year = {2025},
month = {apr},
publisher = {EDP Sciences, IOP Publishing and Società Italiana di Fisica},
volume = {149},
number = {6},
pages = {61001},
author = {Goerlich, Rémi and Hoek, Laura and Chor, Omer and Rahav, Saar and Roichman, Yael},
title = {Experimental realizations of information engines: Beyond proof of concept},
journal = {Europhys. Lett.}
}

@article{garcia_optimal_2025,
  title = {Optimal Closed-Loop Control of Active Particles and a Minimal Information Engine},
  author = {Garcia-Millan, Rosalba and Sch\"uttler, Janik and Cates, Michael E. and Loos, Sarah A. M.},
  journal = {Phys. Rev. Lett.},
  volume = {135},
  issue = {8},
  pages = {088301},
  numpages = {7},
  year = {2025},
  month = {Aug},
  publisher = {American Physical Society},
  doi = {10.1103/fbgp-qpvv},
  url = {https://link.aps.org/doi/10.1103/fbgp-qpvv}
}

@article{sanders_minimal_2025,
  title = {Minimal work protocols for inertial particles in nonharmonic traps},
  author = {Sanders, Julia and Baldovin, Marco and Muratore-Ginanneschi, Paolo},
  journal = {Phys. Rev. E},
  volume = {111},
  issue = {3},
  pages = {034127},
  numpages = {9},
  year = {2025},
  month = {Mar},
  publisher = {American Physical Society},
  doi = {10.1103/PhysRevE.111.034127},
  url = {https://link.aps.org/doi/10.1103/PhysRevE.111.034127}
}

@article{sanders_optimal_2024,
  author       = {Julia Sanders and Marco Baldovin and Paolo Muratore‑Ginanneschi},
  title        = {Optimal Control of Underdamped Systems: An Analytic Approach},
  journal      = {J. Stat. Phys.},
  volume       = {191},
  number       = {9},
  pages        = {117},
  year         = {2024},
  month        = sep,
  publisher    = {Springer Nature},
  doi          = {10.1007/s10955-024-03320-w},
  url          = {https://link.springer.com/10.1007/s10955-024-03320-w}
}

@article{archambault_information_2025,
  title = {Information Engine Fueled by First-Passage Times},
  author = {Archambault, Aubin and Crauste-Thibierge, Caroline and Imparato, Alberto and Jarzynski, Christopher and Ciliberto, Sergio and Bellon, Ludovic},
  journal = {Phys. Rev. Lett.},
  volume = {135},
  issue = {14},
  pages = {147101},
  numpages = {8},
  year = {2025},
  month = {Oct},
  publisher = {American Physical Society},
  doi = {10.1103/s9kj-lczm},
  url = {https://link.aps.org/doi/10.1103/s9kj-lczm}
}

@article{archambault_inertial_2024,
doi = {10.1209/0295-5075/ad8bf0},
url = {https://dx.doi.org/10.1209/0295-5075/ad8bf0},
year = {2024},
month = {nov},
publisher = {EDP Sciences, IOP Publishing and Società Italiana di Fisica},
volume = {148},
number = {4},
pages = {41002},
author = {Archambault, Aubin and Crauste-Thibierge, Caroline and Ciliberto, Sergio and Bellon, Ludovic},
title = {Inertial effects in discrete sampling information engines},
journal = {Europhys. Lett.}
}

@book{seifert_stochastic_2025, 
    place={Cambridge}, 
    title={Frontmatter}, 
    booktitle={Stochastic Thermodynamics}, 
    publisher={Cambridge University Press}, 
    author={Seifert, Udo}, 
    year={2025},
    url = {https://www.cambridge.org/core/books/stochastic-thermodynamics/1766FFBF10FCC7A75DAA89C6E09ED0AB}
}

@article{louwerse_information_2022,
  title = {Information Thermodynamics of the Transition-Path Ensemble},
  author = {Louwerse, Miranda D. and Sivak, David A.},
  journal = {Phys. Rev. Lett.},
  volume = {128},
  issue = {17},
  pages = {170602},
  numpages = {6},
  year = {2022},
  month = {Apr},
  publisher = {American Physical Society},
  doi = {10.1103/PhysRevLett.128.170602},
  url = {https://link.aps.org/doi/10.1103/PhysRevLett.128.170602}
}

@article{paneru_reaching_2020,
  title = {Reaching and violating thermodynamic uncertainty bounds in information engines},
  author = {Paneru, Govind and Dutta, Sandipan and Tlusty, Tsvi and Pak, Hyuk Kyu},
  journal = {Phys. Rev. E},
  volume = {102},
  issue = {3},
  pages = {032126},
  numpages = {6},
  year = {2020},
  month = {Sep},
  publisher = {American Physical Society},
  doi = {10.1103/PhysRevE.102.032126},
  url = {https://link.aps.org/doi/10.1103/PhysRevE.102.032126}
}

@article{barros_probabilistic_2024,
  title = {Probabilistic Work Extraction on a Classical Oscillator Beyond the Second Law},
  author = {Barros, Nicolas and Ciliberto, Sergio and Bellon, Ludovic},
  journal = {Phys. Rev. Lett.},
  volume = {133},
  issue = {5},
  pages = {057101},
  numpages = {5},
  year = {2024},
  month = {Jul},
  publisher = {American Physical Society},
  doi = {10.1103/PhysRevLett.133.057101},
  url = {https://link.aps.org/doi/10.1103/PhysRevLett.133.057101}
}

@incollection{ciliberto_landauers_2021,
	address = {Cham},
	title = {{Landauer's} {Bound} and {{Maxwell}'s} {Demon}},
	isbn = {978-3-030-81480-9},
	url = {https://doi.org/10.1007/978-3-030-81480-9_3},
	booktitle = {Information {Theory}: {Poincaré} {Seminar} 2018},
	publisher = {Springer International Publishing},
	author = {Ciliberto, Sergio},
	editor = {Duplantier, Bertrand and Rivasseau, Vincent},
	year = {2021},
	doi = {10.1007/978-3-030-81480-9_3},
	pages = {87--112},
}

@article{paneru_optimal_2018,
  title = {Optimal tuning of a {Brownian} information engine operating in a nonequilibrium steady state},
  author = {Paneru, Govind and Lee, Dong Yun and Park, Jong-Min and Park, Jin Tae and Noh, Jae Dong and Pak, Hyuk Kyu},
  journal = {Phys. Rev. E},
  volume = {98},
  issue = {5},
  pages = {052119},
  numpages = {7},
  year = {2018},
  month = {Nov},
  publisher = {American Physical Society},
  doi = {10.1103/PhysRevE.98.052119},
  url = {https://link.aps.org/doi/10.1103/PhysRevE.98.052119}
}

@article{paneru_2022_colossal,
  author = {Govind Paneru and Sandipan Dutta and Hyuk Kyu Pak},
  title = {Colossal Power Extraction from Active Cyclic {Brownian} Information Engines},
  journal = {J. Phys. Chem. Lett.},
  volume = {13},
  number = {24},
  pages = {6912--6918},
  year = {2022},
  doi = {10.1021/acs.jpclett.2c01736},
  url = {https://doi.org/10.1021/acs.jpclett.2c01736}
}

@article{ma_reaction_2025,
	title = {Reaction {Coordinates} {Are} {Optimal} {Channels} of {Energy} {Flow}},
	volume = {76},
	issn = {1545-1593},
	url = {https://www.annualreviews.org/content/journals/10.1146/annurev-physchem-082423-010652},
	doi = {https://doi.org/10.1146/annurev-physchem-082423-010652},
	number = {Volume 76, 2025},
	journal = {Annu. Rev. Phys. Chem.	},
	author = {Ma, Ao and Li, Huiyu},
	year = {2025},
	pages = {153--179},
}

@article{lee_brownian_2020,
  title={{Brownian} heat engine with active reservoirs},
  author={Lee, Jaesung and Park, Jae-Mok and Park, Hyunggyu},
  journal={Phys. Rev. E},
  volume={102},
  number={3},
  pages={032116},
  year={2020},
  doi={10.1103/PhysRevE.102.032116},
  url={https://doi.org/10.1103/PhysRevE.102.032116}
}

@article{datta_second_2022,
  title={Second law for active heat engines},
  author={Datta, Amit and Pietzonka, Patrick and Barato, Andre C.},
  journal={Phys. Rev. X},
  volume={12},
  number={3},
  pages={031034},
  year={2022},
  doi={10.1103/PhysRevX.12.031034},
  url={https://doi.org/10.1103/PhysRevX.12.031034}
}

@article{cocconi_efficiency_2024,
  title = {Efficiency of an autonomous, dynamic information engine operating on a single active particle},
  author = {Cocconi, Luca and Chen, Letian},
  journal = {Phys. Rev. E},
  volume = {110},
  issue = {1},
  pages = {014602},
  numpages = {9},
  year = {2024},
  month = {Jul},
  publisher = {American Physical Society},
  doi = {10.1103/PhysRevE.110.014602},
  url = {https://link.aps.org/doi/10.1103/PhysRevE.110.014602}
}

@article{malgaretti_szilard_2022,
  title={{Szilard} engines and information-based work extraction for active systems},
  author={Malgaretti, Paolo and Stark, Holger},
  journal={Phys. Rev. Lett.},
  volume={129},
  pages={228005},
  year={2022},
  doi={10.1103/PhysRevLett.129.228005},
  url={https://doi.org/10.1103/PhysRevLett.129.228005}
}

@article{kim_entropy_2004,
  title={Entropy production of {Brownian} macromolecules with inertia},
  author={Kim, Kyung Hyuk and Qian, Hong},
  journal={Phys. Rev. Lett.},
  volume={93},
  number={12},
  pages={120602},
  year={2004},
  doi={10.1103/PhysRevLett.93.120602},
  url={https://doi.org/10.1103/PhysRevLett.93.120602}
}

@book{bechhoefer_control_2021,
  title={Control Theory for Physicists},
  author={Bechhoefer, John},
  year={2021},
  publisher={Cambridge University Press},
  address={Cambridge, United Kingdom},
  url={https://www.cambridge.org/core/books/control-theory-for-physicists/21AFE5D6C475D1B44BCF9B8536338D98}
}

@article{dago_adiabatic_2023,
  title={Adiabatic computing for optimal thermodynamic efficiency of information processing},
  author={Dago, Santiago and Ciliberto, Sergio and Bellon, Ludovic},
  journal={Proc. Natl. Acad. Sci. U.S.A.},
  volume={120},
  pages={e2301742120},
  year={2023},
  doi={10.1073/pnas.2301742120},
  url={https://doi.org/10.1073/pnas.2301742120}
}

@article{dago_dynamics_2022,
  title={Dynamics of information erasure and extension of {Landauer’s} bound to fast processes},
  author={Dago, Santiago and Bellon, Ludovic},
  journal={Phys. Rev. Lett.},
  volume={128},
  pages={070604},
  year={2022},
  doi={10.1103/PhysRevLett.128.070604},
  url={https://doi.org/10.1103/PhysRevLett.128.070604}
}

@article{dago_information_2021,
  title={Information and thermodynamics: fast and precise approach to {Landauer’s} bound in an underdamped micromechanical oscillator},
  author={Dago, Santiago and Pereda, Javier and Barros, Nestor and others},
  journal={Phys. Rev. Lett.},
  volume={126},
  pages={170601},
  year={2021},
  doi={10.1103/PhysRevLett.126.170601},
  url={https://doi.org/10.1103/PhysRevLett.126.170601}
}

@article{still_thermodynamic_2020,
  title={Thermodynamic cost and benefit of memory},
  author={Still, Susanne},
  journal={Phys. Rev. Lett.},
  volume={124},
  number={5},
  pages={050601},
  year={2020},
  doi={10.1103/PhysRevLett.124.050601},
  url={https://doi.org/10.1103/PhysRevLett.124.050601}
}

@article{paneru_lossless_2018,
  title={Lossless {Brownian} information engine},
  author={Paneru, Govind and Lee, Dong Yun and Tlusty, Tsvi and Pak, Hyung Kyu},
  journal={Phys. Rev. Lett.},
  volume={120},
  number={2},
  pages={020601},
  year={2018},
  doi={10.1103/PhysRevLett.120.020601},
  url={https://doi.org/10.1103/PhysRevLett.120.020601}
}

@article{admon_experimental_2018,
  title={Experimental realization of an information machine with tunable temporal correlations},
  author={Admon, Tamir and Rahav, Saar and Roichman, Yael},
  journal={Phys. Rev. Lett.},
  volume={121},
  number={18},
  pages={180601},
  year={2018},
  doi={10.1103/PhysRevLett.121.180601},
  url={https://doi.org/10.1103/PhysRevLett.121.180601}
}

@article{ribezzi-crivellari_large_2019,
  title={Large work extraction and the {Landauer} limit in a continuous {Maxwell} demon},
  author={Ribezzi-Crivellari, Marco and Ritort, Felix},
  journal={Nat. Phys.},
  volume={15},
  pages={660--664},
  year={2019},
  publisher={Nature Publishing Group},
  doi={10.1038/s41567-019-0472-6},
  url={https://doi.org/10.1038/s41567-019-0472-6}
}

@book{bender_advanced_1978,
  title     = {Advanced Mathematical Methods for Scientists and Engineers},
  author    = {Bender, Carl M. and Orszag, Steven A.},
  year      = {1978},
  publisher = {McGraw-Hill},
  url       = {https://link.springer.com/book/10.1007/978-1-4757-3069-2}
}

@article{leighton_flow_2025,
  author    = {M. P. Leighton and D. A. Sivak},
  title     = {Flow of Energy and Information in Molecular Machines},
  journal   = {Annu. Rev. Phys. Chem.},
  volume    = {76},
  pages     = {379–403},
  year      = {2025},
  note      = {Review article; arXiv preprint available},
  doi      = {https://doi.org/10.1146/annurev-physchem-082423-030023}
}

@article{leighton_arbitrage_2024,
  author    = {M. P. Leighton and J. Ehrich and D. A. Sivak},
  title     = {Information Arbitrage in Bipartite Heat Engines},
  journal   = {Phys. Rev. X},
  volume    = {14},
  pages     = {041038},
  year      = {2024},
  doi       = {10.1103/PhysRevX.14.041038}
}

@article{tsuruyama_rna_2023,
  author  = {T. Tsuruyama},
  title   = {RNA polymerase is a unique {Maxwell’s} demon that converts its transcribing genetic information to free energy for its movement},
  journal = {Eur. Phys. J. Plus},
  volume  = {138},
  pages   = {1},
  year    = {2023},
  doi      = {https://doi.org/10.1140/epjp/s13360-023-04191-y}
}

@article{leighton_efficiencies_2023,
  author  = {M. P. Leighton and D. A. Sivak},
  title   = {Inferring subsystem efficiencies in bipartite molecular machines},
  journal = {Phys. Rev. Lett.},
  volume  = {130},
  pages   = {178401},
  year    = {2023},
  doi      = {https://doi.org/10.1103/PhysRevLett.130.178401}
}

@article{parrondo_thermodynamics_2015,
  author  = {J. M. R. Parrondo and J. M. Horowitz and T. Sagawa},
  title   = {Thermodynamics of information},
  journal = {Nature Phys.},
  volume  = {11},
  pages   = {131--139},
  year    = {2015},
  doi      = {https://doi.org/10.1038/nphys3230}
}

@article{lucero_fluctuation_2021,
  author  = {J. N. E. Lucero and J. Ehrich and J. Bechhoefer and others},
  title   = {Maximal fluctuation exploitation in Gaussian information engines},
  journal = {Phys. Rev. E},
  volume  = {104},
  pages   = {044122},
  year    = {2021},
  doi     = {10.1103/PhysRevE.104.044122}
}

@book{kloeden_numerical_1992,
	address = {Berlin, Heidelberg},
	title = {Numerical {Solution} of {Stochastic} {Differential} {Equations}},
	copyright = {http://www.springer.com/tdm},
	isbn = {978-3-642-08107-1 978-3-662-12616-5},
	url = {http://link.springer.com/10.1007/978-3-662-12616-5},
	urldate = {2025-02-02},
	publisher = {Springer Berlin Heidelberg},
	author = {Kloeden, Peter E. and Platen, Eckhard},
	year = {1992},
	doi = {10.1007/978-3-662-12616-5},
}

@inproceedings{saha_optical_2021,
  author    = {T. K. Saha and J. Bechhoefer},
  title     = {Optical trapping and optical micromanipulation},
  booktitle = {Proc. SPIE Int. Soc. Opt. Eng.	},
  volume    = {11798},
  pages     = {53--61},
  publisher = {SPIE},
  year      = {2021},
  doi      = {https://doi.org/10.1117/12.2593992}
}

@article{saha_bayesian_2022,
  author  = {T. K. Saha and J. N. E. Lucero and J. Ehrich and others},
  title   = {Bayesian information engine that optimally exploits noisy measurements},
  journal = {Phys. Rev. Lett.},
  volume  = {129},
  pages   = {130601},
  year    = {2022},
  doi     = {10.1103/PhysRevLett.129.130601}
}

@article{saha_nonequilibrium_2023,
  author  = {T. K. Saha and J. Ehrich and M. Gavrilov and others},
  title   = {Information engine in a nonequilibrium bath},
  journal = {Phys. Rev. Lett.},
  volume  = {131},
  pages   = {057101},
  year    = {2023},
  doi     = {10.1103/PhysRevLett.131.057101}
}

@article{du_buisson_performance_2024,
	title = {Performance limits of information engines},
	volume = {9},
	issn = {null},
	url = {https://www.tandfonline.com/doi/full/10.1080/23746149.2024.2352112},
	doi = {10.1080/23746149.2024.2352112},
	number = {1},
	urldate = {2025-02-02},
	journal = {Adv. Phys.: X},
	author = {du Buisson, Johan and Sivak, David A. and Bechhoefer, John},
	month = dec,
	year = {2024},
	note = {Publisher: Taylor \& Francis},
	pages = {2352112},
}

@article{li_fundamental_2012,
  author    = {Tongcang Li and Simon Kheifets and Mark G. Raizen},
  title     = {Millikelvin cooling of an optically trapped microsphere in vacuum},
  journal   = {Nat. Phys.},
  volume    = {7},
  number    = {7},
  pages     = {527--530},
  year      = {2011},
  doi       = {10.1038/nphys1952},
  url       = {https://doi.org/10.1038/nphys1952}
}

@article{gieseler_feedback_2012,
  author  = {J. Gieseler and B. Deutsch and R. Quidant and others},
  title   = {Subkelvin parametric feedback cooling of a laser-trapped nanoparticle},
  journal = {Phys. Rev. Lett.},
  volume  = {109},
  pages   = {103603},
  year    = {2012},
  doi     = {10.1103/PhysRevLett.109.103603}
}

@article{tebbenjohanns_cold_2019,
  author  = {F. Tebbenjohanns and M. Frimmer and A. Militaru and others},
  title   = {Cold damping of an optically levitated nanoparticle to microkelvin temperatures},
  journal = {Phys. Rev. Lett.},
  volume  = {122},
  pages   = {223601},
  year    = {2019},
  doi     = {10.1103/PhysRevLett.122.223601}
}

@article{serreli_molecular_2007,
  author  = {V. Serreli and C. Lee and E. R. Kay and others},
  title   = {A molecular information ratchet},
  journal = {Nature},
  volume  = {445},
  pages   = {523--527},
  year    = {2007},
  doi     = {10.1038/nature05452}
}

@article{berut_verification_2012,
  author  = {A. B{\'e}rut and others},
  title   = {Experimental verification of {Landauer’s} principle linking information and thermodynamics},
  journal = {Nature},
  volume  = {483},
  pages   = {187--189},
  year    = {2012},
  doi     = {https://doi.org/10.1038/nature10872}
}

@article{jun_test_2014,
  author  = {Y. Jun and M. Gavrilov and J. Bechhoefer},
  title   = {High-precision test of {Landauer’s} principle in a feedback trap},
  journal = {Phys. Rev. Lett.},
  volume  = {113},
  pages   = {190601},
  year    = {2014},
  doi      = {https://doi.org/10.1103/PhysRevLett.113.190601}
}

@article{koski_mutual_2014,
  author  = {J. V. Koski and V. F. Maisi and T. Sagawa and J. P. Pekola},
  title   = {Experimental observation of the role of mutual information in the nonequilibrium dynamics of a {Maxwell} demon},
  journal = {Phys. Rev. Lett.},
  volume  = {113},
  pages   = {030601},
  year    = {2014},
  doi      = {https://doi.org/10.1103/PhysRevLett.113.030601}
}

@article{hong_test_2016,
  author  = {J. Hong and B. Lambson and D. Scott and J. Bokor},
  title   = {Experimental test of {Landauer’s} principle in single-bit operations on nanomagnetic memory bits},
  journal = {Sci. Adv.},
  volume  = {2},
  pages   = {e1501492},
  year    = {2016},
  doi      = {https://doi.org/10.1126/sciadv.1501492}
}

@article{toyabe_information_2010,
  author  = {S. Toyabe and T. Sagawa and M. Ueda and E. Muneyuki and M. Sano},
  title   = {Experimental demonstration of information-to-energy conversion and validation of the generalized {Jarzynski} equality},
  journal = {Nat. Phys.},
  volume  = {6},
  pages   = {988--992},
  year    = {2010},
  doi     = {10.1038/nphys1821}
}

@article{camati_entropy_2016,
  author  = {P. A. Camati and others},
  title   = {Experimental {rectification} of {entropy} {production} by {Maxwell’s} {demon} in a {quantum} {system}},
  journal = {Phys. Rev. Lett.},
  volume  = {117},
  pages   = {240502},
  year    = {2016},
  doi     = {10.1103/PhysRevLett.117.240502}
}

@article{koski_refrigerator_2015,
  author  = {J. V. Koski and A. Kutvonen and I. M. Khaymovich and T. Ala-Nissila and J. P. Pekola},
  title   = {On-chip {Maxwell’s} demon as an information-powered refrigerator},
  journal = {Phys. Rev. Lett.},
  volume  = {115},
  pages   = {260602},
  year    = {2015},
  doi     = {10.1103/PhysRevLett.115.260602}
}

@article{cottet_quantum_2017,
  author  = {N. Cottet and others},
  title   = {Observing a quantum {Maxwell} demon at work},
  journal = {Proc. Natl. Acad. Sci. U.S.A.},
  volume  = {114},
  pages   = {7561--7564},
  year    = {2017},
  doi     = {https://doi.org/10.1073/pnas.1704827114}
}

@article{masuyama_conversion_2018,
  author  = {Y. Masuyama and others},
  title   = {Information-to-work conversion by {Maxwell’s} demon in a superconducting circuit quantum electrodynamical system},
  journal = {Nat. Commun.},
  volume  = {9},
  pages   = {1291},
  year    = {2018},
  doi     = {https://doi.org/10.1038/s41467-018-03686-y}
}

@article{koski_szilard_2014,
  author  = {J. V. Koski and V. F. Maisi and J. P. Pekola and D. V. Averin},
  title   = {Experimental realization of a {Szilard} engine with a single electron},
  journal = {Proc. Natl. Acad. Sci. U.S.A.},
  volume  = {111},
  pages   = {13786--13789},
  year    = {2014},
  doi      = {https://doi.org/10.1073/pnas.1406966111}
}

@article{chida_power_2017,
  author  = {K. Chida and S. Desai and K. Nishiguchi and A. Fujiwara},
  title   = {Power generator driven by {Maxwell’s} demon},
  journal = {Nat. Commun.},
  volume  = {8},
  pages   = {15310},
  year    = {2017},
  doi     = {https://doi.org/10.1038/ncomms15301}
}

@article{vandenbroeck_ensemble_2015,
  author  = {C. Van den Broeck and M. Esposito},
  title   = {Ensemble and trajectory thermodynamics: A brief introduction},
  journal = {Physica A},
  volume  = {418},
  pages   = {6--16},
  year    = {2015},
  doi     = {10.1016/j.physa.2014.04.035}
}

@article{seifert_stochastic_2012,
  author  = {U. Seifert},
  title   = {Stochastic thermodynamics, fluctuation theorems and molecular machines},
  journal = {Rep. Prog. Phys.},
  volume  = {75},
  pages   = {126001},
  year    = {2012},
  doi     = {10.1088/0034-4885/75/12/126001}
}

@book{sekimoto_stochastic_2010,
  author    = {K. Sekimoto},
  title     = {Stochastic Energetics},
  publisher = {Springer},
  year      = {2010},
  volume    = {799},
  series    = {Lecture Notes in Physics},
  doi       = {10.1007/978-3-642-05411-2}
}

@article{sekimoto_kinetic_1997,
  author  = {K. Sekimoto},
  title   = {Kinetic characterization of heat bath and the energetics of thermal ratchet models},
  journal = {J. Phys. Soc. Jpn.},
  volume  = {66},
  pages   = {1234--1237},
  year    = {1997},
  doi     = {10.1143/JPSJ.66.1234}
}

@article{bennett_thermodynamics_1982,
  author  = {C. H. Bennett},
  title   = {The thermodynamics of computation---A review},
  journal = {Int. J. Theor. Phys.},
  volume  = {21},
  pages   = {905--940},
  year    = {1982},
  doi     = {10.1007/BF02084158}
}

@article{landauer_irreversibility_1961,
  author  = {R. Landauer},
  title   = {Irreversibility and heat generation in the computing process},
  journal = {IBM J. Res. Dev.},
  volume  = {5},
  pages   = {183--191},
  year    = {1961},
  doi     = {10.1147/rd.53.0183}
}

@article{szilard_maxwell_2003,
  author    = {Leo Szilard},
  title     = {On the Decrease of Entropy in a Thermodynamic System by the Intervention of Intelligent Beings},
  journal   = {Behav. Sci.},
  volume    = {9},
  number    = {4},
  pages     = {301--310},
  year      = {1964},
  doi       = {10.1002/bs.3830090402},
  publisher = {Wiley},
  url       = {https://onlinelibrary.wiley.com/doi/10.1002/bs.3830090402}
}

@article{szilard_entropie_1929,
  author  = {L. Szilard},
  title   = {{\"U}ber die Entropieverminderung in einem thermodynamischen System bei Eingriffen intelligenter Wesen},
  journal = {Z. Angew. Phys.},
  volume  = {53},
  pages   = {840--856},
  year    = {1929},
  doi     = {10.1007/BF01341281},
  url     = {https://link.springer.com/article/10.1007/BF01341281}
}

@book{knot_tait_1911,
  author    = {C. G. Knott},
  title     = {Life and Scientific Work of Peter Guthrie Tait},
  publisher = {Cambridge University Press},
  year      = {1911},
  address   = {London},
  url = {https://archive.org/details/lifescientificwo00knotuoft}
}

@article{lee_experimentally-achieved_2018,
	title = {An experimentally-achieved information-driven {Brownian} motor shows maximum power at the relaxation time},
	volume = {8},
	issn = {2045-2322},
	url = {https://www.nature.com/articles/s41598-018-30495-6},
	doi = {10.1038/s41598-018-30495-6},
	number = {1},
	urldate = {2025-04-08},
	journal = {Scientific Reports},
	author = {Lee, Dong Yun and Um, Jaegon and Paneru, Govind and Pak, Hyuk Kyu},
	month = aug,
	year = {2018},
	pages = {12121},
}

@article{
saha_maximizing_2021,
author = {Tushar K. Saha  and Joseph N. E. Lucero  and Jannik Ehrich  and David A. Sivak  and John Bechhoefer},
title = {Maximizing power and velocity of an information engine},
journal = {Proc. Natl. Acad. Sci. U.S.A.},
volume = {118},
number = {20},
pages = {e2023356118},
year = {2021},
url = {https://www.pnas.org/doi/abs/10.1073/pnas.2023356118}
}

\end{document}